\documentclass{article}
\usepackage{graphicx} % Required for inserting images
\usepackage{amsmath} % For math-related commands
\usepackage{amsfonts}
\usepackage{array}
\usepackage{algorithm}
\usepackage{algpseudocode}
\usepackage{bm}
\usepackage{natbib}
\usepackage{booktabs}

\newcommand{\Tao}{\mathcal{T}}

\usepackage{authblk}
\title{Geometric Deep Learning for Realized Covariance Matrix Forecasting}
\author[1]{Andrea Bucci}
\affil[1]{Department of Economics and Law, University of Macerata, Italy}
\author[2]{Michele Palma}
\affil[2]{Department of Computing Science, Bocconi University, Italy}
\author[3]{Chao Zhang}
\affil[3]{Financial Technology Thrust, The Hong Kong University of Science and Technology, Guangzhou, China}

\date{}

\begin{document}

\maketitle

\begin{abstract}
Traditional methods employed in matrix volatility forecasting often overlook the inherent Riemannian manifold structure of symmetric positive definite matrices, treating them as elements of Euclidean space, which can lead to suboptimal predictive performance. Moreover, they often struggle to handle high-dimensional matrices. In this paper, we propose a novel approach for forecasting realized covariance matrices of asset returns using a Riemannian-geometry-aware deep learning framework. In this way, we account for the geometric properties of the covariance matrices, including possible non-linear dynamics and efficient handling of high-dimensionality. Moreover, building upon a Fréchet sample mean of realized covariance matrices, we are able to extend the HAR model to the matrix-variate. We demonstrate the efficacy of our approach using daily realized covariance matrices for the 50 most capitalized companies in the S\&P 500 index, showing that our method outperforms traditional approaches in terms of predictive accuracy.
\end{abstract}

\normalsize

\section{Introduction}

In finance, the role of the predicted volatility of asset returns is crucial for portfolio optimization and risk management \citep{ma52}. However, forecasting covariance matrices can be challenging since the predicted matrix must be symmetric positive definite (SPD). In GARCH models, this has been ensured either through a quadratic form in the BEKK model \citep{engkro95}, by separately modeling variances and correlations as in the Dynamic Conditional Correlation (DCC) model \citep{en02}, or through factor models \citep{Lanne2007, Vrontos2003}.

The introduction of a nonparametric way to estimate volatility, \textit{i.e.} the realized volatility \citep{abdl01, Andersen2001, Andersen2003}, has opened the way for new methods to ensure that the matrix is SPD. This issue has been addressed by parametrizing the realized covariance (RCOV) matrices using techniques like Cholesky decomposition \citep{chivo11} or the matrix logarithm function \citep{bavo11, Asai2022}, followed by the application of time series models on the transformed data. However, the use of appropriate transformations of the covariance matrix makes the direct interpretation of parameters difficult, since the time series models are applied to the non-linearly transformed series \citep{bucci2022comparing}. Moreover, for very large matrices, methods based on diagonalization, such as the matrix logarithm transformation, may not be feasible due to computational limitations, as large matrices are difficult to diagonalize. For these reasons, other strands of the multivariate volatility literature have instead focused on the use of time-varying Wishart distributions to model the dynamics of the whole covariance matrix \citep{Jin2016, Yu2017} and the combination of GARCH models with realized covariances \citep{nosheshe11}.  Nevertheless, all these methods still suffer from the proliferation of parameters when the number of assets is high. Some methods have been proposed to overcome this limitation, such as a multi-step procedure, the use of the least absolute shrinkage and selection operator (LASSO) regression \citep{Callot2017}, or the application of a sparse structure for the covariance matrix \citep{Asai2015, Gribisch2020}. More recently, \cite{Opschoor2024} proposed to use the Conditional Autoregressive $F$-Riesz Model to model the matrix-variate time series of realized covariances. The problem of managing high-dimensional covariance matrices still remains, and there is room for proposing innovative methods. 

An additional fundamental limitation of all these methods is that they treat SPD matrices as if they were elements of the Euclidean space, overlooking the fact that SPD matrices inherently rely on a Riemannian manifold - a type of curved geometric space equipped with a specific metric that locally resembles the Euclidean space \citep{marron2022}. Covariance matrices,  as specific examples of SPD matrices, indeed form a Riemannian manifold known as the manifold of symmetric positive definite matrices. Merely applying traditional Euclidean methods results in inadequate for the purpose of forecasting these matrices, because this may lead to suboptimal predictive performance or convergence failures of estimation methods when dealing with high-dimensional matrices. In order to address the aforementioned issues, it is crucial to employ tools from differentiable geometry as accounting for the geometry of SPD matrices has been shown to both enhance predictive power and improve the handling of parameter dimensionality \citep{harandi2014manifold, tuzel2008pedestrian, huang2017riemannian}. Early attempts to leverage the Riemannian geometry of SPD matrices resulted in the development of the Riemannian distance learning methods \citep{zhao2023machine, li2012electroencephalogram}. Within the framework of predicting multivariate volatility using the geometric properties of the covariance matrices, the literature has provided limited research. For instance, \cite{Han2022} suggested using a geometric covariance dynamics framework in which the intrinsic geometric properties of covariance matrices are specified through differential geometry. Still, they do not account for possible nonlinearities in the dynamics of the covariances. To overcome this limitation, \cite{Lee2024} recently proposed to consider a regime-switching GARCH-like structure on the Riemannian SPD manifold. 

A possible choice to somehow consider both nonlinear dynamics and the geometric properties of RCOV is to use geometric deep learning techniques, in particular those designed to operate in SPD manifolds \citep{chakraborty2018statistical, brooks2019riemannian, wang2022dreamnet}. A notable example is the Riemannian Neural Network for SPD Matrix Learning (SPDNet) introduced by \cite{huang2017riemannian}, which allows for non-linearly learning of a more desirable SPD matrix by directly optimizing on manifolds, starting from a high-dimensional input SPD matrix.  
A deep learning geometrically-aware approach for the forecasting of realized covariance matrices, with improved performances and designed to efficiently handle high-dimensional data, could boost multivariate financial volatility forecasting. However, existing networks have primarily been designed for classification tasks and continue to rely on traditional Euclidean layers in the final output stage to make predictions.
Moreover, since these networks are designed primarily for dimensionality reduction of the input matrix, they are structured to process a single input matrix during the learning process. This approach limits their applicability for the time series regression task, where multiple lags are involved. 
In this work, we introduce a novel application of Riemannian neural networks for forecasting RCOV matrices. We extend the SPDNet architecture to adapt it for regression purposes, by also enabling multiple input SPD matrices to be allowed in the learning process. 

In the context of our task, this approach facilitates the usage of multiple lagged covariance matrices as predictors for the matrix to be predicted.
We accomplish this by constructing the input as a diagonal block matrix of lagged covariance matrices. This simple way of including multiple lags in our input matrix allows us to easily account for volatility time series persistence \citep{Baillie1996, Hurvich2005} also in high-dimensional problems, which, for instance, is not possible in methods which combine a vector autoregressive (VAR) model and a parametrization without using a factor model. This is particularly relevant as numerous studies have reported the presence of long memory in financial volatility \citep{Andersen2003, Gatheral2018}. By incorporating up to 10 lagged covariance matrices, our approach effectively captures the persistence in volatility series, which is not feasible in other adopted models due to the prohibitive increase in the number of parameters. In this context, our approach can also be used to extend the Heterogeneous Autoregressive (HAR) model introduced by \cite{cor09} to the entire covariance matrix, leveraging the Fréchet mean of the covariance matrices to aggregate information at the weekly and monthly levels. By constructing an input matrix that integrates the daily lagged RCOV, the weekly average, and the monthly average realized covariance matrices into a block-diagonal form, we maintain the geometric properties essential for accurate forecasting. This method not only preserves the SPD nature of the matrices but also mitigates the issue of parameter explosion inherent in traditional multivariate extensions of the HAR model. In addition, the sparsity introduced by the diagonal block matrix allows us to reduce computational time and somehow avoid local minima in the Riemannian optimization problem.

%%Say also about using a proper loss function for covariance matrices, which could be positive semi-definite with deficient rank

 In summary, our contribution is threefold: i) we introduce a way to predict realized covariance matrices of asset returns based on a deep neural network, that preserves the geometric structure of SPD matrices without relying on any arbitrary parametrization; ii) we introduce the diagonal block matrix construction approach to allow SPDNet to handle multiple SPD matrices as input and to account for the well-known long-memory characteristic of volatilities; iii) we extend to the matrix-variate framework the HAR model of \cite{cor09}, making it possible to use such a framework also in a high-dimensional context.

We test our approach on daily realized covariance matrices of the fifty most capitalized companies comprising the S\&P 500 index.  A comparison with some traditional methods highlights that the approaches introduced in this paper that account for long-term persistence of volatility give more accurate predictions of RCOV than alternative models.

The remainder of the paper is organized as follows. In Section \ref{sec:RCOV}, we introduce the geometric properties of realized covariance matrices of asset returns along with a brief introduction to fundamental concepts related to manifolds and their role in optimization problems. This shall provide the reader with a more comprehensive understanding of the inner mechanisms of Riemannian neural networks. In Section \ref{sec:SPDNet}, we briefly recall the SPDNet architecture and present our contributions to it, including the geometric version of the HAR, while the statistical comparison of the forecasts on real data is presented in Section \ref{sec:Empirical}. The predictions are then used in a portfolio optimization application in Section \ref{sec:Portfolio}. Section \ref{sec:Conclusions} concludes.

\section{Realized Covariance Matrices as Riemannian manifolds} \label{sec:RCOV}
The realized covariance matrices of asset returns are constructed from observed prices at high frequencies. The construction in the general setting occurs as follows. Suppose we have intraday prices for $n$ different stocks, from which it is possible to derive the high-frequency return as
\begin{equation*}
   r_{i,\tau} = \displaystyle\ln\left({\frac{P_{i,\tau}}{P_{i,\tau-1}}}\right),
\end{equation*}
where the subscript $i$ is referred to the $i$-th asset at the intraday period $\tau$. Gathering in a vector all the data about intraday returns of the assets $\mathbf{r}_\tau = [r_{1,\tau}, \ldots, r_{n,\tau}]^\top$, we can use it to build the realized covariance matrix at day $t$ as the sum of cross products of all returns of a given day $t$ as follows \citep{Andersen2003, bashe02}:
\begin{equation}\label{eq:RC}
    \mathbf{RC}_t = \displaystyle\sum_{\tau=1}^{N_t} \mathbf{r}_{\tau} \mathbf{r}_{\tau}^\top,
\end{equation}
where $N_t$ is the number of intraday observations in the $t$-th day. It must be noticed that such realized covariance matrices $\mathbf{RC}_t$, $t=1, \ldots, T$, are SPD.
% \begin{enumerate}
%     \item $\mathbf{RC}_t$ is symmetric:  $$\mathbf{RC}_t = \mathbf{RC}_t^\top;$$
%     \item $\mathbf{RC}_t$ is positive definite: \begin{equation}
%    \mathbf{x}^\top \mathbf{RC}_t \mathbf{x} > 0 \quad \text{for} \quad \mathbf{x} \neq \mathbf{0}.
%  \end{equation}
%  \end{enumerate}
The collection of all symmetric positive definite matrices of a given size lies on a special kind of topological space known as a \textit{smooth manifold}, where each point possesses a tangent space that locally resembles Euclidean space. When equipped with a specific metric, referred to as a Riemannian metric, this space becomes a Riemannian manifold. 
It is possible to enable effective manipulation of covariance matrices while preserving their intrinsic geometric properties by using tools of Riemannian geometry.
Furthermore, it is possible to develop inference techniques that rely on the underlying manifold structure, and related works have shown that this approach can lead to better predictions \citep{huang2017riemannian}.

\subsection{Riemannian geometry}
In the following sections, we will provide a non-rigorous introduction to the fundamentals of Riemannian geometry. This overview aims to convey the essential concepts necessary for understanding the framework and context of this work. While the discussion will be simplified for clarity, it is intended to equip the reader with a foundational grasp of Riemannian geometry to facilitate comprehension of the subsequent analyses and discussions.
To introduce manifolds and subsequently Riemannian manifolds, the related fundamental concept of topological space needs to be recalled.

A \textit{topological space} is a set $X$ with a topology $\Tao  \subseteq \mathcal{P}(X)$ defined on it. 
The topology allows for the notion of closeness between elements of $X$ to be defined without necessarily relying on a numeric distance measure.
To achieve this, the following properties must be satisfied:
\begin{itemize}
    \item both $\emptyset $ and $X$ belong to $\Tao$;
    \item the union of any collection of elements of $\Tao$ belongs to $\Tao$;
    \item the intersection of any finite number of elements of $\Tao$ belongs to $\Tao$.
\end{itemize}
The collection $\Tao$ is known as a topology on $X$ and its elements are referred to as open sets. A specific topological space with an additional structure can be found in manifolds. An $n$-dimensional \textit{manifold}, $M$, is a topological space that locally resembles the $n$-dimensional Euclidean space near each point. The notion of \textit{atlas} underlies the formal definition of manifold, before defining it, the notion of \textit{chart} shall be introduced.

A \textit{chart} for $M$ is a pair $(U, \varphi)$ where $U \subseteq M$ is an open set and $\varphi\colon U \rightarrow V$, known as \textit{coordinate map}, is a homeomorphism onto an open set $V \subseteq \mathbb{R}^n$.

An \textit{atlas} for $M$ is an indexed family of charts on M, recorded as $$\{(U_{\alpha},\varphi_{\alpha})\colon \alpha \in I\}$$ such that it covers $M$, that is $$\bigcup \limits_{\alpha}^{} U_{\alpha} = M.$$
Overlapping between charts may occur, this motivates the need for a \textit{transition map}, which describes how the coordinates in one chart relate to the coordinates of the other. 
Two charts $(U_{\alpha},\varphi_{\alpha}) \text{ and } (U_{\beta},\varphi_{\beta})$ s.t. $U_{\alpha} \cap U_{\beta} \neq \emptyset$ with transition maps $\varphi_{\beta} \circ \varphi_{\alpha}^{-1}\colon \varphi_{\alpha}(U_{\alpha}\cap U_{\beta}) \rightarrow \varphi_{\beta}(U_{\alpha}\cap U_{\beta})$ and $\varphi_{\alpha} \circ \varphi_{\beta}^{-1}\colon \varphi_{\beta}(U_{\alpha}\cap U_{\beta}) \rightarrow \varphi_{\alpha}(U_{\alpha}\cap U_{\beta})$ 
are said to be \textit{compatible} if their transition maps are differentiable. Recall that transition maps are defined as homeomorphism between open subsets of $\mathbb{R}^n$. If compatibility holds for every pair of overlapping charts in an atlas, the atlas is differentiable and $M$ is said to be a \textit{differentiable manifold}.
When transition functions are infinitely differentiable $C^{\infty}$-maps, then the atlas is called a smooth atlas and the manifold itself is said to be a \textit{smooth manifold}.
The concepts of tangent vectors and directional derivatives can be clearly defined on a differentiable manifold; for the purposes of the following introductory work, let $M$ be a differentiable manifold.

The \textit{tangent space} at a point $p \in M$ denoted with $T_pM$, is a vector space whose elements are all possible tangent vectors at $p$, which intuitively represent the possible directions in which one can infinitesimally move from $p$ while remaining on the manifold. 
Such space is a first-order approximation of $M$ around $p$, and its dimension at every point of $M$ is the same as the dimension of the manifold itself.

In order to allow for the measurement of angles, lengths, and distances on the manifold, a family of smoothly varying inner products on the tangent space at each point of $M$ must be defined.
If for any point $p \in M$ holds that $g_p\colon (T_pM \times T_pM) \rightarrow \mathbb{R}$ associates to each $p$ a smooth varying positive definite symmetric bilinear form on $T_pM$, then the metric $g$ that assigns the inner product $g_p$, is a Riemannian metric. 
The smooth manifold $M$ equipped with Riemannian metric $g$ forms a \textit{Riemannian manifold}, denoted with the pair $(M,g)$. We shall simplify the terminology when referring to Riemannian manifolds by using only $M$ and leaving the Riemannian metric implicit.

A particular Riemannian manifold that will be central to the SPDNet architecture is the Stiefel manifold $St(n,k)$, a set of orthonormal $k$-frames in $\mathbb{R}^n$, that is
\begin{equation}
     St(n,k) = \{x \in \mathbb{R}^{k\times n} | x^Tx = \mathbb{I}_{n} \}.
\end{equation}
An orthonormal $k$-frame is a collection of $k$ orthonormal vectors in an $n$-dimensional space.

In the following section, we shall introduce some optimization concepts related to Riemannian manifolds that will prove useful in addressing our work.

\subsection{Riemannian Optimization}
It is possible to define a smooth function $f\colon M \rightarrow \mathbb{R}$, and consequentially  define a class of optimization problems in the form: \begin{align*}
    \min \quad & f(x) \\
    \text{subject to} \quad & x \in M
    \end{align*}
where the search space is constrained to a Riemannian manifold, $M$. Direct application of gradient-based optimization algorithms in Euclidean space, such as Stochastic Gradient Descent (SGD), proves ineffective for optimization problems constrained by the geometry of the Symmetric Positive Definite (SPD) manifold. However, these algorithms can be readily adapted to Riemannian settings to address such optimization challenges. The crucial point of such translation is the computation of Riemannian gradient  $\tilde{\nabla}f(x)$, which is obtained as orthogonal projection, $\pi_x[\nabla f(x)]$, of Euclidean gradient onto tangent space $T_xM$. It must be observed that the Euclidean gradient $\nabla f(x)$ is a vector in the ambient space $\mathbb{R}^n$ and it does not necessarily lie in $T_xM$, this supports the need for a Riemannian gradient that takes into account the curvature of $M$ when used in the updating step.
Before proceeding further, the notions of \textit{exponential map} and \textit{logarithm map} will be presented.

The \textit{exponential map}, $exp_p \colon T_pM \rightarrow M$ at point $p \in M$ moves vectors from tangent space to a point on a manifold, \textit{i.e.} for a given vector $v \in T_pM$, the point $u = \exp_p(v)$ is obtained by moving along geodesic that starts at $p$ in direction of $v$ for a unit time.

The \textit{logarithm map} at a given point $p \in M$ is a map defined as follows, $\log_p\colon  M \rightarrow T_pM$, if $exp_p(v) = u$ then $\log_p(u)=v$, where geodesics are curves in compliance with the manifold geometry that locally minimize the distance between two points of $M$. They can be thought of as a generalization of straight lines to curved spaces.

It must be noticed that the exact form of the projection $ \pi_x \colon\mathbb{R}^n \rightarrow T_xM$ depends on how $M$ is embedded in $\mathbb{R}^n$. For some manifolds, like for the Stiefel manifold, the exact form is known, and specifically, it is obtained as a subtraction of the normal component of the Euclidean gradient. Below, we present the Riemannian Gradient Descent (RGD) algorithm to solve the introduced Riemannian optimization problem.

\begin{algorithm}
    \caption{Riemannian Gradient Descent}
    \begin{algorithmic}[1]
    \State \textbf{Input:} Initial point $x_0 \in M$, learning rate $\eta > 0$
    \State Initialize $x \gets x_0$
    \While{not converged}
        \State Compute Euclidean Gradient $\nabla f(x)$
        \State Compute Riemannian Gradient $\tilde{\nabla} f(x) = \pi_x(\nabla f(x))$
        \State Update: $
        x \gets \exp_{x}(-\eta \tilde{\nabla}f(x)) $
    \EndWhile
    \State \textbf{Return} $x$
    \end{algorithmic}
    \end{algorithm}
    Since the computation of the exponential map $\exp_{x}(v)$ may entail significant computational complexity, it is often replaced by retraction map $R_x(v)$ which is a first-order approximation of the exponential map. The exact form  of the map varies based on the specification of the manifold and ambient space, for Stiefel manifold it has the form of orthogonal projection, $R_x(v) = orth(v)$. To guarantee that output results in an orthonormal matrix, techniques like Singular Value Decomposition (SVD) or QR decomposition can be used.
%maybe summarize this section%

\section{SPDNet for regression} \label{sec:SPDNet}
 
The concepts presented in the previous section also apply to the realized covariance matrices, which, as stated before, lie on a Riemannian manifold. In this paper, we predict the realized covariance matrices relying on a Riemannian Neural Network, which can guarantee that the predicted outputs are still a point on the SPD manifold, avoiding any kind of direct parametrization of matrices to guarantee the SPD property and accounting also for the nonlinear behavior of volatility dynamics \citep{sch89}.

Specifically, we adapt the SPD matrix learning network (SPDNet) proposed by \citep{huang2017riemannian} to regression problems. The ReSPDNet introduced in this paper designs two types of layers: a fully connected convolution-like layers (bilinear mapping, BiMap), and a rectified linear units (ReLU)-like layers for the eigenvalue rectification (ReEig). The BiMap layers aim to convert the input Symmetric Positive Definite (SPD) matrices into new SPD matrices using a bilinear mapping, without any parametrization. Instead, the proposed ReEig layer brings non-linearity to the SPDNet, as it rectifies the resultant SPD matrix using a non-linear function, specifically the ReLU activation function.

The BiMap layer transforms the input SPD matrix to new SPD matrices as follows:
\begin{equation}
 \mathbf{X}_k = f_k(\mathbf{X}_{k-1};\mathbf{W}_k) = \mathbf{W}_k\mathbf{X}_{k-1}\mathbf{W}_k^\top,   
 \label{eq:bimap}
\end{equation}
where $\mathbf{X}_{k-1}$ is the input matrix, of size say $d_{k-1}  \times d_{k-1}$, coming from the previous layer, $\mathbf{X}_k$ is the output matrix of size $d_k \times d_k$  and $\mathbf{W}_k$ is the weight matrix of size $ d_k \times d_{k-1}$. The dimensionality of $\mathbf{X}_k$ shall be equal to or lower than the input $\mathbf{X}_{k-1}$, however, we can also expand the input matrix, whenever $d_k\geq d_{k-1}$, as follows:

$$\mathbf{Z} = \mathbf{A} + \mathbb{I}\cdot \mathbf{X_{k-1}} \cdot \mathbb{I}^\top$$
where $\mathbf{A}$ is a diagonal matrix of size $d_k \times d_k$, the first $n$ diagonal elements are zeros, and remaining $d_k-d_{k-1}$ elements are ones, and $\mathbb{I}$ is a matrix of size $d_k \times d_{k-1}$ that resembles an identity matrix on the first $d_{k-1}$ rows, and has zero elements on entries of remaining $d_k-d_{k-1}$ rows. Then, we apply the BiMap layer considering the matrix $\mathbf{Z}$ as the input matrix for the layer, following equation \eqref{eq:bimap}. This possibility of expanding the matrix is also supported by the empirical findings presented in the relevant section of our work.

The ReEig ($k$-th) layer rectifies the input SPD matrix for the current layer, by enhancing their small positive eigenvalues, obtaining a new SPD matrix $X_k$ as follows:
$$\mathbf{X}_k = f_k(\mathbf{X}_{k-1})= \mathbf{U}_{k-1} \max (\epsilon \mathbf{I},\mathbf{\Sigma}_{k-1})\mathbf{U}_{k-1}^\top$$ where  $\mathbf{I}$ is the identity matrix, $\epsilon$ is the rectification threshold, and $\mathbf{U}_{k-1}$, $\mathbf{\Sigma}_{k-1}$ are obtained by eigenvalue decomposition of $\mathbf{X}_{k-1}$.

%The LogEig layer is used to perform Riemannian computing on resulting SPD matrices for output layers with objective functions. We exclude this layer as we are not interested in classification tasks, indeed the final output of our network will be an SPD matrix, not a class prediction. 

It must be noticed that, in both cases, $\mathbf{W}_k$ is initialized such that it is semi-orthogonal. This latter requirement comes from the following observation: to ensure $\mathbf{X}_k$ results in an SPD matrix, $\mathbf{W}_k$ must be row full-rank.
Unfortunately, finding such $\mathbf{W}_k$ that minimizes loss function results to be a Riemannian optimization problem on a non-compact Stiefel manifold, and a direct optimization is practically infeasible. 
To overcome such an issue, we additionally require $\mathbf{W}_k$ to be semi-orthogonal such that the weights will lie on compact Stiefel manifold, $ St(d_k, d_{k-1})$. To solve this Riemannian optimization problem, the Riemannian gradient descent (specifically, Riemannian stochastic gradient descent) for Stiefel Manifolds is used.
Recall that the steepest descent direction for the corresponding loss function, in a given layer $k$, with respect to our weights $\mathbf{W}_k$ on the Stiefel manifold, has to be found in the Riemannian gradient \citep{huang2017riemannian}. Progressing along this tangential direction results in updates within the tangent space of the Stiefel manifold. Subsequently, this update undergoes mapping back to the Stiefel manifold through a retraction operation. 

Having presented the architecture of the network to be utilized, we will now address the challenge of constructing the input matrix to accommodate multiple lagged covariance matrices. This matrix, which will be fed into the first layer of our network —a BiMap layer— must be designed to ensure that the network's output is an SPD matrix of the same size as the single covariance matrix that we aim to forecast.

For further details on the implementation of stochastic gradient descent for Stiefel Manifold, and the backpropagation algorithm used to compute the Euclidean gradient in the presence of the layers mentioned above, please refer to \cite{huang2017riemannian}.

\subsection{Constructing the input matrix}
Let $n$ be the number of stocks, our dataset will be a time series of RCOV matrices, $\mathbf{RC}_1, ..., \mathbf{RC}_T$, each of size $n\times n$. Let $k$ be the number of input lags to be used, the resulting final dataset will be defined as follows, $\{(\mathbf{D}_t,\mathbf{Y}_t)\mid \mathbf{D}_t \in I,  \mathbf{Y}_t \in O \}$ where 
 $O = \{\mathbf{RC}_k, ...,\mathbf{RC}_T\}$ and $I=\{\mathbf{D}_k,...,\mathbf{D}_T\}$, where each $\mathbf{D}_t$ is obtained as a diagonal block matrix of $k$-input lagged matrices preceding $\mathbf{Y}_t$:
\begin{equation}\label{eq:diagonal}
    \mathbf{D}_t = \begin{bmatrix}
\mathbf{Y}_{t-1} & 0 & 0 & \cdots & 0 \\
0 & \mathbf{Y}_{t-2} & 0 & \cdots & 0 \\
0 & 0 & \mathbf{Y}_{t-3} & \cdots & 0 \\
\vdots & \vdots & \vdots & \ddots & \vdots \\
0 & 0 & 0 & \cdots & \mathbf{Y}_{t-k}
\end{bmatrix}.
\end{equation}

It must be noticed that the construction of a block diagonal matrix using SPD matrices is still an SPD matrix. Furthermore, $\mathbf{D}_t$ will not have size $n \times n$ but SPDNET will reduce the input SPD matrix to a desired output matrix of size $n \times n$, still preserving SPD property.

At each training step, $\mathbf{D}_t$ will be the input for the first layer of our network. In this regard, we provide in Section \ref{sec:Empirical} the description of the tested network architectures.

\subsection{Geometric HAR (GeoHAR)}
The Heterogeneous Autoregressive (HAR) model introduced by \citet{cor09} has been a cornerstone in forecasting realized volatility (RV) given its ability to capture long memory within a simple framework. By incorporating realized volatilities over different time horizons—daily, weekly, and monthly—the HAR model accounts for the persistence and heterogeneity observed in financial volatility dynamics. However, when extending the HAR model to a multivariate context, existing approaches often stack the unique elements of the covariance matrix in a vector. This vectorization leads to a proliferation of parameters in time series models used to make predictions, mining the computational feasibility of the estimation and the efficiency of the estimates, especially as the number of assets increases.

In this section, we propose a geometric extension of the HAR model to the entire covariance matrix. 
Our geometric HAR (GeoHAR) model extends the original framework proposed by \citet{cor09} to operate directly on SPD manifolds using the ReSPDNet introduced before. By doing so, we capture the temporal dependencies and nonlinear dynamics of the entire covariance matrix without relying on arbitrary parameterizations or excessive computational overhead. This innovative approach enhances forecasting accuracy and efficiency, particularly in high-dimensional settings common in financial applications. To do so, we employ the Fréchet mean of covariance matrices to aggregate information across different time horizons while respecting the manifold structure of symmetric positive definite (SPD) matrices.

Consider a probability distribution for a $n \times n$ realized covariance matrix, $\mathbf{S} = \mathbf{RC}$, on a Riemannian metric space with density $f(\mathbf{S})$ \citep{Dryden2009}. The Fr\'{e}chet mean, $\bm{\Sigma}$, is defined as
\begin{equation}
	\bm{\Sigma} = \arg \underset{\bm{\Sigma}}{\inf}\frac{1}{2}\int d(\mathbf{S}, \bm{\Sigma})^2 f(\mathbf{S})d\mathbf{S},
\end{equation}
where $d(\cdot)$ is a proper distance for covariance matrices \citep{Dryden2009}. With a sample $\mathbf{S}_1, \ldots, \mathbf{S}_T$ of observations, if they are naively imposed to be i.i.d., the sample Fr\'{e}chet mean is computed as
\begin{equation}\label{eq:samplefrec}
\overline{\bm{\Sigma}} = \arg \underset{\bm{\Sigma}}{\inf} \sum_{t= 1}^{T}d(\mathbf{S}_t, \bm{\Sigma})^2.
\end{equation}
In our approach, we first specify $d(\cdot)$ as a log-Euclidean distance \citep{Arsigny2007} defined as follows
\begin{equation}\label{eq:riem}
	d(\mathbf{S}_1, \mathbf{S}_2) = ||\text{logm}(\mathbf{S}_1)-\text{logm}(\mathbf{S}_2)||_F,
\end{equation}
where logm$(\cdot)$ is the matrix logarithm function of the argument. This leads to the following Fréchet sample mean
\begin{equation}
\overline{\bm{\Sigma}} = \text{expm}\left\{\arg \underset{\bm{\Sigma}}{\inf} \sum_{t= 1}^{T}||\text{logm}(\mathbf{S}_1)-\text{logm}(\bm{\Sigma})||_F\right\} = \text{expm} \left\{\frac{1}{T}\sum_{t=1}^{T}    \text{logm}(\mathbf{S}_t)\right\},
\end{equation}
where expm$\{\cdot\}$ denotes the exponential matrix function of the argument.

It should be noted that the log-Euclidean distance cannot deal with positive semi-definite matrices exhibiting rank deficiency, which may happen in the case of high-dimensional realized covariance matrices \citep{Reiss2021}. Nevertheless, this limitation can be overcome by projecting $\mathbf{S}_t$ in the space of SPD matrices using an ad hoc approximant, such as the one proposed by \cite{Higham1988}. Alternatively, we can use a non-Euclidean size-and-shape metric defined as
\begin{equation}
    d(\mathbf{S}_1, \mathbf{S}_2) = \arg \underset{\mathbf{R}\in O(n)}{\inf} ||\mathbf{L}_1 - \mathbf{L}_2\mathbf{R}||,
\end{equation}
where $\mathbf{L}_i$ is a decomposition of $\mathbf{S}_i$, such that $\mathbf{S}_i = \mathbf{L}_i\mathbf{L}_i^\top$, for $i = 1, 2$. The Procrustes solution for matching $\mathbf{L}_2$ and $\mathbf{L}_1$ is
\begin{equation}\label{eq:proc}
    \hat{\mathbf{R}} = \arg \underset{\mathbf{R}\in O(n)}{\inf} ||\mathbf{L}_1 - \mathbf{L}_2\mathbf{R}|| = \mathbf{UW}^\top,
\end{equation}
where $\mathbf{L}^\top \mathbf{L} = \mathbf{W}\bm{\Lambda}\mathbf{U}^\top$, for $\mathbf{U, W} \in O(n)$ and $\bm{\Lambda}$ is a diagonal matrix of positive singular values \citep{Dryden2009}. If we have a sample $\mathbf{S}_1, \ldots, \mathbf{S}_T$ observations, the sample Fréchet mean is computed as in Equation \eqref{eq:samplefrec}, which, in dual size-and-shape formulation, can be written as 
\begin{equation*}
    \overline{\bm{\Sigma}} = \hat{\bm{\Delta}}\hat{\bm{\Delta}}^\top,
\end{equation*}
where
\begin{equation*}
    \hat{\bm{\Delta}} =  \arg \underset{\bm{\Delta}}{\inf}  \sum_{t=1}^T \underset{\mathbf{R}\in O(n)}{\inf} ||\mathbf{H}^\top \mathbf{L}_i\mathbf{R}_i - \mathbf{H}^\top \bm{\Delta}||^2,
\end{equation*}
whose solution can be obtained using the Generalized Procrustes Algorithm \citep{Gower1975}.

By computing the Fréchet mean over the previous 5 and 22 observations—mirroring the weekly and monthly horizons in the original HAR model—we obtain two $n \times n$ sample mean of covariance matrices
\begin{align*}
    \overline{\mathbf{RC}}_{t-1}^w &= \arg \underset{\mathbf{RC}}{\inf} \sum_{i= 1}^{5}d(\mathbf{RC}_{t-i}, \mathbf{RC})^2,
\\
    \overline{\mathbf{RC}}_{t-1}^m &=  \arg \underset{\mathbf{RC}}{\inf} \sum_{i= 1}^{22}d(\mathbf{RC}_{t-i}, \mathbf{RC})^2,
\end{align*}
which can be used to construct the input diagonal matrix in a HAR-like fashion as follows: 
\begin{equation}\label{eq:Dthar} 
\mathbf{D}_t = \begin{bmatrix} \mathbf{RC}_{t-1}^d & 0 & 0 \\ 0 & \overline{\mathbf{RC}}_{t-1}^w & 0\\ 0 & 0 & \overline{\mathbf{RC}}_{t-1}^m \end{bmatrix}, \end{equation} 
where $\mathbf{RC}_{t-1}^d$ is the previous lag and the suffix \lq d\rq \ only denotes the daily nature of the memory component.
In this way, we extend the HAR model to the matrix-variate framework and we capture the dynamics of the entire covariance matrix across multiple time scales within a geometric context.

%An alternative way to build $\mathbf{D}_i$ is to use the repeated application (pairwise) of the Kronecker product. Both methods guarantee for an input matrix of size bigger than the resulting output matrix, indeed the network return as prediction SPD matrix of size same size as the single original observations.

\section{Empirical application}\label{sec:Empirical}

\subsection{Data}
The empirical application of our forecasting method utilizes time series data comprising intradaily close spot prices for the 50 largest U.S. stocks by market capitalization, as of November 2021, all of which are part of the S\&P 500 index. This data obtained from Lobster spans the period from June 27th, 2007 to July 1st, 2021. Table \ref{table:tickers} provides a comprehensive list of the stock names and their corresponding tickers. Daily realized volatility matrices were computed according to Equation \eqref{eq:RC}, resulting in a total of $T=3409$ daily observations. The covariance stationarity of each time series was validated through unit root tests, including the Augmented Dickey-Fuller (ADF), ADF-GLS, Phillips-Perron (PP), and KPSS tests. %Figure \ref{fig:plots} illustrates the realized volatilities for the $n = 20$ stocks.

\begin{table}[htbp]
	\centering
	\caption{\textbf{List of the 50 US stocks used in this study}}\label{table:tickers}
	\begin{tabular}{p{0.2\textwidth}p{0.2\textwidth}p{0.2\textwidth}p{0.2\textwidth}}
		\hline
		\textbf{Ticker} & \textbf{Stock}&\textbf{Ticker} & \textbf{Stock}\\
		\hline
AAPL & Apple Inc. & JPM & JPMorgan Chase \& Co.\\
ABT  & Abbott Laboratories & KO & Coca-Cola Co.\\
ACN  & Accenture & LLY& Eli Lilly \& Co.\\
ADBE & Adobe Inc. & LOW& Lowe's Companies Inc.\\
AMGN & Amgen Inc. & MA & Mastercard Inc.\\
AMT  & American Tower Corp. & MCD& McDonald's Corp.\\
AMZN & Amazon.com Inc. & MRK & Merck \& Co. Inc.\\
AXP  & American Express Co. & MSFT& Microsoft Corp.\\
BAC  & Bank of America Corp. & NFLX & Netflix Inc.\\
CAT  & Caterpillar Inc. & NKE & Nike Inc.\\
CMCSA& Comcast Corp. & NVDA& NVIDIA Corp.\\
COST & Costco Wholesale Corp.& ORCL & Oracle Corp.\\
CRM  & Salesforce Inc. & PEP & PepsiCo, Inc.\\
CSCO & Cisco Systems Inc. & PFE & Pfizer Inc.\\
CVX  & Chevron Corp. & PG  & Procter \& Gamble Co.\\
DHR  & Danaher Corp. & QCOM & Qualcomm Inc.\\
DIS  & The Walt Disney Co. & T & AT\&T Inc.\\
GE   & General Electric Co. & TJX& TJX Companies Inc.\\
GOOG &	Alphabet Inc. Class A. & TMO & Thermo Fisher Scientific Inc.\\
GS   & Goldman Sachs Group Inc. & TXN & Texas Instruments Inc.\\
HD   &  Home Depot Inc.&  UNH & UnitedHealth Group Inc.\\
IBM  & IBM Corp. & UNP & Union Pacific Corp.\\
INTU & Intuit Inc. & VZ & Verizon Communications Inc.\\
ISRG & Intuitive Surgical Inc. & WFC & Wells Fargo \& Co.\\
JNJ  &  Johnson \& Johnson &  WMT & Walmart Inc.\\
\hline
\end{tabular}
\end{table}

\subsection{Network specifications}
We have evaluated two distinct network configurations, drawing inspiration from the ones proposed by \citep{huang2017riemannian}. We employ two BiMap layers along with two RegEig layers. The layers are arranged such that each BiMap layer at level $k$ is immediately followed by a RegEig layer at level $k+1$, but the final output layer always consists of a BiMap layer. This arrangement is crucial for enhancing predictive power, in the specific case of RCOV matrices, given the particular behaviour of the ReLU activation function to clip all negative values.
Using a single lag, the block diagonal matrix, that is the input of the network will result in a $50 \times 50$ matrix, indeed no block diagonal matrix is constructed. Whereas, using 3 lags will result in an input matrix of size $150 \times 150$, and so on.  In general, the size of the block diagonal matrix of $k$ lagged RCOV matrices, each of size $n \times n$, will result in a matrix of size $kn \times kn$. For the Geometric version of the HAR, the size of the input matrix is always $150 \times 150$.
%We used 125/75 as intermediate layers

\subsection{Forecasting results}
We produce out-of-sample predictions from the SPDNet and we compare them to forecasts from a Factorial VAR(1) model applied to the parameter of a Cholesky decomposition \citep{chivo11} stacking the returns in alphabetical order, using 50 principal components. 
We also include in the set of competing models the GO-GARCH model \citep{vanderWeide2002} and a random walk prediction (\textit{i.e.}, $\widehat{\mathbf{RC}}_{t+1} = \mathbf{RC}_t$). When we use a parametrization, we model and predict the parameters of the realized covariance matrices, then we construct back the positive semidefinite prediction of RCOV. The one-step-ahead predictions are obtained using a rolling window method with a window of $2386$ observations, this means that we have $T_{test}=1023$ out-of-sample predictions.

To assess the predictive accuracy of the methods under comparison, we employ the Frobenius, Euclidean and Procrustes distances between the observed and predicted matrices as both metrics and loss functions. According to \cite{Laurent2013}, the first two distances demonstrate robustness in relation to volatility proxies. Instead, the Procrustes distance can be applied to rank-deficient covariance matrices, which is a common feature in high-dimensional RCOVs \citep{Reiss2021, bucci2022comparing}.
These metrics are both used as averages to assess the best-performing method in terms of predictive accuracy, and as inputs for the Model Confidence Set (MCS) test \citep{haluna11}. This method relies on an equivalence test, denoted as $\delta_\mathcal{M}$, and an elimination rule, $e_{\mathcal{M}}$. The process begins by applying the equivalence test to the initial set of models, $\mathcal{M} = \mathcal{M}_0$, where $\mathcal{M}_0$ represents the complete set of competing models. If the equivalence test $\delta_\mathcal{M}$ is rejected, it indicates that the models in $\mathcal{M}_0$ are not equally effective. Consequently, the elimination rule $e_{\mathcal{M}}$ is employed to remove the model that exhibits the poorest performance based on the sample data. This procedure is iteratively repeated until the equivalence test $\delta_\mathcal{M}$ is no longer rejected, at which point the remaining models constitute the Set of Superior Models (SSM). In this study, we utilize the range statistic, $T_R$, as the test statistic for the Model Confidence Set (MCS) procedure, which is calculated as follows:
\begin{equation*}
T_R = \underset{i,j \in \mathcal{M}}{\max} \mid t_{ij}\mid, \quad i,j \in \mathcal{M},
\end{equation*}
where
\begin{equation*}
t_{ij} = \frac{\bar{d}_{ij}}{\sqrt{\widehat{\text{var}}(\bar{d}_{ij})}}.
\end{equation*}
Here, $\bar{d}_{ij}$ represents the average loss difference between models $i$ and $j$, calculated as $\bar{d}_{ij} = T^{-1}\sum_{t=1}^{T}(L_{i,t} - L_{j,t})$, where $L_{i,t}$ is the loss associated with model $i$ at time $t$. The term $\widehat{\text{var}}(\bar{d}_{ij})$ denotes the estimated variance of $\bar{d}_{ij}$.

The forecasting results in terms of predictive accuracy are reported in Table \ref{table:Losses20}. In the table, the \texttt{ReSPDNET} models are trained either using the mean squared error or log-Euclidean loss functions. The subscripts \lq 1\rq, \lq 3\rq, \lq 5\rq \ and \lq 10\rq \ identify the number of lags used to build the input matrix, while the subscripts \lq LE\rq \ and \lq P\rq \ denote the use of either log-Euclidean or Procrustes distance to compute the Fréchet mean in the GeoHAR models. \texttt{Cholesky} refers to the Cholesky decomposition, while \texttt{RW} concerns predictions from a random walk model. The $p$-value refers to the probability of being included in the SSM over 10000 block bootstrap replicates. 

It can be noticed that the lowest average Frobenius is obtained by the ReSPDNet with five lags and trained using the log-Euclidean distance, closely followed by the same architecture but with 3 and 10 lags respectively. Moreover, the only models entering the SSM are the Geometric HAR models trained using MSE, the one that combines a FAVAR with Cholesky decomposition, and the ReSPDNet-LE trained with 10 lags. We then use as a statistical measure the Euclidean distance, since we want to understand if even using a more \lq traditional\rq \ loss function, our approach is still capable of outperforming the alternatives.  Once again, the best-performing model is the SPDNet trained with a log-Euclidean loss and 10 lags, which is also the only model in the SSM, strictly followed by the same model with 5 lags. If we observe the last two columns of Table \ref{table:Losses20} concerning the use of the Procrustes metric, we can state that the best-performing model is still the ReSPDNET-LE$_{10}$, but also that the Geometric HAR relying on a log-Euclidean distance used both as a loss function and to compute the Fréchet mean has a very similar performance. 

These results underline the relevance of a proper geometric-aware loss function for training and evaluating a predictive model for covariance matrices. In fact, in terms of average loss functions, the ReSPDNETs trained using MSE are always outperformed by those obtained using a log-Euclidean distance for the Riemannian gradient. Moreover, this way of inserting lags into the model, even if it may seem naive, allows us to account for longer memory that, even with a factorial model, would be unfeasible with traditional methods which are typically estimated using a single lag. Finally, even if the GeoHARs do not significantly outperform the models with 10 lags, the computational time for their training is severely lower, and this could be a good compromise between traditional and more complex models.

\begin{table}[htbp] 
	\caption{\textbf{Average loss functions and MCS $p$-values for $n=50$ assets}}\label{table:Losses20}
	\begin{tabular}{l>{\raggedleft\arraybackslash}p{1.8cm}@{\hskip 0.2cm}r@{\hskip 0.2cm}>{\raggedleft\arraybackslash}p{1.8cm}@{\hskip 0.3cm}r@{\hskip 0.2cm}>{\raggedleft\arraybackslash}p{1.8cm}@{\hskip 0.3cm}r}
		\toprule
		& \multicolumn{2}{c}{\textbf{Frobenius}} & \multicolumn{2}{c}{\textbf{Euclidean}}& \multicolumn{2}{c}{\textbf{Procrustes}}\\
		\cmidrule(lr){2-3} \cmidrule(lr){4-5} \cmidrule(lr){6-7}% spazio tra le \cmidrule
		\textbf{Model} 	& \textbf{Avg. loss}   &  $p_{\textsc{MCS}}$    & \textbf{Avg. loss}       &  $p_{\textsc{MCS}}$     & \textbf{Avg. loss}       &  $p_{\textsc{MCS}}$   \\
		\midrule
		\texttt{ReSPNET-LE$_{1}$}  & 3937.07            &0.220         & 25.394               &$<$0.001         & 3.042            & $<$0.001\\
		\texttt{ReSPNET-LE$_{3}$}  & 2907.51            &0.597         & 21.329               &$<$0.001         & 2.343            & $<$0.001\\
		\texttt{ReSPNET-LE$_{5}$}  & \underline{2813.32}&\textbf{1.000}& 20.429               &0.726          & 2.183            & 0.618\\
		\texttt{ReSPNET-LE$_{10}$} & 3040.69            &\textbf{1.000}& \underline{20.215}   &\textbf{1.000} & \underline{2.174}& \textbf{1.000}\\[5 pt]
        \texttt{ReSPNET-M$_{1}$}   & 5208.35            &0.278         & 28.380               &$<$0.001         & 3.628            & $<$0.001\\
		\texttt{ReSPNET-M$_{3}$}   & 4114.38            &0.597         & 23.947               &$<$0.001          & 2.831            & $<$0.001\\
		\texttt{ReSPNET-M$_{5}$}   & 4029.64            &0.597         & 23.549               &0.001          & 2.742            & $<$0.001\\
		\texttt{ReSPNET-M$_{10}$}  & 4065.22            &0.597         & 23.691               &$<$0.001          & 2.753            & $<$0.001\\[5 pt]
		\texttt{GeoHAR-LE$_{LE}$}  & 4246.43            &0.597         & 21.904               &0.111          & 2.178            & \textbf{0.990}\\
		\texttt{GeoHAR-LE$_{P}$}   & 3198.60            &0.597         & 21.788               &$<$0.001          & 2.321            & $<$0.001\\[5 pt]
        \texttt{GeoHAR-M$_{LE}$}   & 8445.82            &\textbf{1.000}& 25.010               &0.197          & 2.816            & 0.209\\
		\texttt{GeoHAR-M$_{P}$}    & 9347.37            &\textbf{0.849}& 26.436               &0.191          & 2.992            & 0.618\\[5 pt]
		\texttt{Cholesky}          & 3570.12            &\textbf{0.826}& 21.100               &0.044          & 2.219            & 0.013\\
		\texttt{GO-GARCH}          & 10933.75           &0.140         & 35.906               &$<$0.001          & 2.863            & $<$0.001\\
		\texttt{RW}	               & 3358.24            &0.312         & 22.816               &$<$0.001          & 2.493            & $<$0.001\\
		\bottomrule  
		\multicolumn{7}{p{\textwidth}}{\footnotesize Note: Underlined values denote the lowest loss functions for each setting. Subscripts \lq 1\rq, \lq 3\rq, \lq 5\rq \ and \lq 10\rq \ in the ReSPDNet models identify the number of lags used to build the input matrix, while the subscripts \lq LE\rq \ and \lq P\rq \ in the GeoHAR model denote the use of either log-Euclidean or Procrustes distance to compute the Fréchet mean. The \lq LE\rq \ or \lq M\rq \ at the end of the name of the model (\textit{i.e.}, ReSPDNET-LE) underlines that the model in the row has been trained using either a log-Euclidean or MSE loss function. \texttt{Cholesky} refers to the FAVAR applied to the Cholesky decomposition of RCOV; \texttt{RW} concerns predictions from a random walk model. The $p$-value refers to the probability of being included in the SSM over 10000 block bootstrap replicates (in bold values with a $p_{\textsc{MCS}}$ greater than 0.75).}
	\end{tabular}
\end{table}

\subsection{Performance across stock market volatility regimes}
As a robustness check, we follow the approach of \cite{Zhang2024} and we see how the different models perform throughout market regimes computed as follows: we consider a period of financial distress when the realized variance of the S\&P 500 index is above its 90\% percentile \citep{Pascalau2023, Zhang2024}, a calm period otherwise.

The results related to the calm and turbulent period are reported respectively in Panel A and B of Table \ref{table:LossesRob}. Overall, the findings somehow reflect what is already observed in Table \ref{table:Losses20}. The ReSPDNET with 10 lags is the best-performing method during periods of low and moderate volatility (\textit{i.e.}, Panel A), strictly followed by the GeoHAR which uses the log-Euclidean distance to compute the Fréchet mean. The latter is also the unique method in the SSM when a Procrustes loss function is considered as a metric.

During the turbulent period (\textit{i.e.}, Panel B), the ReSPDNET models with three and five lags seem to outperform all the competing models. Underlining the fact that long memory is a feature of realized covariance matrices but mostly in periods of low volatility. Once again, training a model with a Euclidean loss, \textit{i.e.} the MSE, leads to higher average losses, even if some of the \lq M\rq \ models enter the SSM.

\begin{table}[htbp] 
	\caption{\textbf{Average loss functions and MCS $p$-values for $n=50$ assets}}\label{table:LossesRob}
	\begin{tabular}{l>{\raggedleft\arraybackslash}p{1.8cm}@{\hskip 0.2cm}r@{\hskip 0.2cm}>{\raggedleft\arraybackslash}p{1.8cm}@{\hskip 0.3cm}r@{\hskip 0.2cm}>{\raggedleft\arraybackslash}p{1.8cm}@{\hskip 0.3cm}r}
		\toprule
		& \multicolumn{2}{c}{\textbf{Frobenius}} & \multicolumn{2}{c}{\textbf{Euclidean}}& \multicolumn{2}{c}{\textbf{Procrustes}}\\
		\cmidrule(lr){2-3} \cmidrule(lr){4-5} \cmidrule(lr){6-7}% spazio tra le \cmidrule
		\textbf{Model} 	& \textbf{Avg. loss}   &  $p_{\textsc{MCS}}$    & \textbf{Avg. loss}       &  $p_{\textsc{MCS}}$     & \textbf{Avg. loss}       &  $p_{\textsc{MCS}}$   \\
		\midrule
		\multicolumn{7}{l}{\textbf{Panel A: Calm period}}\\
		\texttt{ReSPNET-LE$_{1}$}  & 811.113            &$<$0.001      & 17.132                  &$<$0.001         & 2.282              & $<$0.001\\
		\texttt{ReSPNET-LE$_{3}$}  & 588.849            &$<$0.001      & 14.256                  &$<$0.001         & 1.767              & $<$0.001\\
		\texttt{ReSPNET-LE$_{5}$}  & 502.538            &0.004         & 13.353                  &$<$0.001         & 1.639              & 0.047\\
		\texttt{ReSPNET-LE$_{10}$} & 454.723            &\textbf{0.888}& \underline{12.773}      &\textbf{1.000}   & 1.624              & 0.158\\[4 pt]
		\texttt{ReSPNET-M$_{1}$}   & 961.637            &$<$0.001      & 18.695                  &$<$0.001         & 2.595              & $<$0.001\\
		\texttt{ReSPNET-M$_{3}$}   & 611.851            &$<$0.001      & 15.154                  &$<$0.001         & 2.072              & $<$0.001\\
		\texttt{ReSPNET-M$_{5}$}   & 608.386            &$<$0.001      & 14.938                  &$<$0.001         & 2.022              & $<$0.001\\
		\texttt{ReSPNET-M$_{10}$}  & 591.287            &$<$0.001      & 14.815                  &$<$0.001         & 2.001              & $<$0.001\\[4 pt]
		\texttt{GeoHAR-LE$_{LE}$}  & \underline{453.293}&\textbf{1.000}& 12.775                  &\textbf{0.989}   & \underline{1.603}  & \textbf{1.000}\\
		\texttt{GeoHAR-LE$_{P}$}   & 617.892            &$<$0.001      & 14.473                  &$<$0.001         & 1.757              & $<$0.001\\[4 pt]
		\texttt{GeoHAR-M$_{LE}$}   & 513.399            &$<$0.001      & 13.816                  &$<$0.001         & 1.832              & $<$0.001\\
		\texttt{GeoHAR-M$_{P}$}    & 566.780            &$<$0.001      & 14.349                  &$<$0.001         & 1.882              & $<$0.001\\[4 pt]
		\texttt{Cholesky}          & 492.536            &0.065         & 13.276                  &$<$0.001         & 1.659              & 0.004\\
		\texttt{GO-GARCH}          & 1136.450           &$<$0.001      & 21.294                  &$<$0.001         & 2.132              & $<$0.001\\
		\texttt{RW}	               & 671.288            &$<$0.001      & 15.398                  &$<$0.001         & 1.914              & $<$0.001\\
		\midrule
		\multicolumn{7}{l}{\textbf{Panel B: Turbulent period}}\\
		\texttt{ReSPNET-LE$_{1}$}  & 31858.23           &0.196         & 99.190             &0.002          & 9.828             & $<$0.001\\
		\texttt{ReSPNET-LE$_{3}$}  & 23617.89           &\textbf{1.000}& 84.510             &\textbf{0.850} & 7.485             & $<$0.001\\
		\texttt{ReSPNET-LE$_{5}$}  &\underline{23453.34}&\textbf{1.000}& \underline{83.636} &\textbf{1.000} & \underline{7.035} & \textbf{1.000}\\
		\texttt{ReSPNET-LE$_{10}$} & 26138.63           &\textbf{1.000}& 86.686             &\textbf{0.764} & 7.083             & \textbf{0.868}\\[4 pt]
		\texttt{ReSPNET-M$_{1}$}   & 43140.19           &0.256         & 114.885            &0.024          & 12.857            & 0.005\\
		\texttt{ReSPNET-M$_{3}$}   & 35399.09           &\textbf{0.808}& 102.487            &0.023          & 9.611             & $<$0.001\\
		\texttt{ReSPNET-M$_{5}$}   & 34588.39           &\textbf{0.808}& 100.466            &0.043          & 9.178             & $<$0.001\\
		\texttt{ReSPNET-M$_{10}$}  & 35094.56           &0.702         & 102.973            &0.045          & 9.465             & 0.004\\[4 pt]
		\texttt{GeoHAR-LE$_{LE}$}  & 38126.86           &0.702         & 103.442            &0.0289         & 7.323             & 0.290\\
		\texttt{GeoHAR-LE$_{P}$}   & 26249.58           &0.482         & 87.125             &0.197          & 7.354             & 0.230\\[4 pt]
		\texttt{GeoHAR-M$_{LE}$}   & 79298.56           &\textbf{1.000}& 124.991            &0.525          & 11.601            & 0.393\\
		\texttt{GeoHAR-M$_{P}$}    & 87775.99           &\textbf{0.891}& 134.392            &\textbf{0.764} & 12.900            & \textbf{0.868}\\[4 pt]
		\texttt{Cholesky}          & 31059.20           &\textbf{0.808}& 90.989             &0.335          & 7.221             & 0.013\\
		\texttt{GO-GARCH}          & 98443.58           &0.281         & 166.417            &$<$0.001       & 9.394             & 0.043\\
		\texttt{RW}	               & 27358.17           &0.701         & 89.069             &0.131          & 7.669             & 0.040\\
		\bottomrule  
		\multicolumn{7}{p{\textwidth}}{\footnotesize Note: Underlined values denote the lowest loss functions for each setting. Subscripts \lq 1\rq, \lq 3\rq, \lq 5\rq \ and \lq 10\rq \ in the ReSPDNet models identify the number of lags used to build the input matrix, while the subscripts \lq LE\rq \ and \lq P\rq \ in the GeoHAR model denote the use of either log-Euclidean or Procrustes distance to compute the Fréchet mean. The \lq LE\rq \ or \lq M\rq \ at the end of the name of the model underlines that the model in the row has been trained using either a log-Euclidean or MSE loss function. \texttt{Cholesky} refers to the FAVAR applied to the Cholesky decomposition of RCOV; \texttt{RW} concerns predictions from a random walk model. The $p$-value refers to the probability of being included in the SSM over 10000 block bootstrap replicates (in bold values with a $p_{\textsc{MCS}}$ greater than 0.75).}
	\end{tabular}
\end{table}

\section{Portfolio optimization}\label{sec:Portfolio}
To evaluate our predictions in terms of economic value \citep{Bollerslev2018}, we compare the performance of portfolios obtained through the Global Minimum Variance (GMV). Assuming a risk-averse investor who wants to invest in $n$ assets using the predicted realized covariance matrix, $\widehat{\mathbf{RC}}_{t+1}$, the GMV weights are obtained from the following optimization problem
\begin{align*}
    &\arg \min  &&\mathbf{w}_{t+1}' \widehat{\mathbf{RC}}_{t+1} \mathbf{w}_{t+1}\\
    & s.t. &&\mathbf{w}_{t+1}'\bm{\iota} = 1
\end{align*}
where $\bm{\iota}$ is a $n \times 1$ vector of ones. In our analysis, $GMV$ indicates a portfolio without no-short-selling constraints. We also compute the optimal portfolio with the long-only constraint, denoted as $GMV^+$, which imposes \textit{i.e.}
$w_{i,t} \geq 0$, $\forall i,t$. Finally, we compute
the portfolio returns as $r_{t,p} = \mathbf{w}_t'\mathbf{r}_t$.

We compare the performance of the portfolios obtained with the same methods used for the statistical comparison in the previous section through the following measures:
\begin{itemize}
    \item Annualized portfolio standard deviation, computed as
    \begin{equation*}
        \sigma_p = \sqrt{252 \cdot \frac{1}{T}\sum_{t=1}^{T}\left(r_{t,p} - \bar{r}_{p}\right)^2};
    \end{equation*}
    \item Average portfolio turnover with adjusted weights
    \begin{align*}
        TO_t &= \sum_{i=1}^N \begin{vmatrix}
            w_{i,t+1} - w_{i,t}\frac{1+r_{i,t}}{1+\mathbf{w}_t' \mathbf{r}_t}
        \end{vmatrix}\\
        \tau_p &= \frac{1}{T-1}\sum_{t=1}^{T-1}TO_t
    \end{align*}
    where $TO_t$ is the portfolio turnover from day $t$ to day $t+1$\footnote{No transaction costs are assumed.}.
\end{itemize}

Table \ref{table:Portfolio} presents the performance metrics of the $GMV$ and $GMV^+$ portfolios constructed using the covariance matrices predicted by different models. We also add the naive equally weighted portfolio with $w_i = 1/n$, since \cite{DeMiguel2009} demonstrate that the naive portfolio is hardly beaten by competing models in terms of turnover. 

Several key aspects emerge. First, traditional parametric frameworks, such as those based on Cholesky decomposition or GO-GARCH specifications, are associated with low annualized standard deviations. Nevertheless, the proposed GeoHAR with Fréchet mean computed using the log-Euclidean distance also exhibits a comparably low standard deviation, demonstrating that geometric deep learning methods show risk levels that are competitive with established benchmark models.

Second, while certain methods excel in risk minimization, they often do so at the expense of higher portfolio turnover. For instance, the \texttt{Cholesky}-based strategy exhibits a substantially elevated turnover, which, in the presence of transaction costs, would likely erode the benefits of lower risk. By contrast, some geometric approaches, such as the \texttt{GeoHAR-LE$_{LE}$}, manage to maintain relatively low turnover while delivering near-optimal risk metrics. This balance highlights the practical appeal of geometric modeling frameworks: they not only preserve the intrinsic structure of the data but also provide stable portfolio allocations that reduce the frequency of rebalancing.

From an investor’s perspective, methods that minimize portfolio variance without inducing excessive trading activity are particularly valuable. The geometric approaches, therefore, present a good compromise, blending model-based risk reduction with operational efficiency. Moreover, the no-short-selling constraint generally increases the standard deviations, but geometric methods, again, demonstrate robustness by sustaining relatively low risk and turnover levels even under these stricter conditions.

\begin{table}[!ht]
	\centering
	\caption{\textbf{Global Minimum Variance Portfolio Performance}}
	\label{table:Portfolio}
	\begin{tabular}{lrr@{\hskip 0.3cm}rr}
		\toprule
		& \multicolumn{2}{c}{GMV} & \multicolumn{2}{c}{GMV$^+$}\\
		\cmidrule(lr){2-3} \cmidrule(lr){4-5}
		&$\sigma_p$ & $\tau_p$ & $\sigma_p$ & $\tau_p$\\
		\midrule
		\texttt{ReSPDNet-LE$_1$}   & 8.920         & 3.248          & 7.354             &1.315 \\
		\texttt{ReSPDNet-LE$_3$}   & 6.619         & 2.648          & 4.841             &1.166  \\
		\texttt{ReSPDNet-LE$_5$}   & 7.012         & 2.282          & 4.523             &1.120 \\
		\texttt{ReSPDNet-LE$_{10}$}& 7.717         & 2.079          & 6.107             &1.161  \\[5 pt]
		\texttt{ReSPDNet-M$_1$}    & 6.389         & 2.717          & 5.735             &1.484 \\
		\texttt{ReSPDNet-M$_3$}    & 7.187         & 3.040          & 6.409             &1.188  \\
		\texttt{ReSPDNet-M$_5$}    & 8.047         & 2.471          & 4.789             &1.004 \\
		\texttt{ReSPDNet-M$_{10}$} & 7.365         & 2.123          & 5.572             &0.938  \\[5 pt]
		\texttt{GeoHAR-LE$_{LE}$}  & \textbf{4.950}& \textbf{1.930} & 4.635             &1.003\\
		\texttt{GeoHAR-LE$_{P}$}   & 7.087         & 2.027          & 6.068             &\textbf{0.924}\\[5 pt]
		\texttt{GeoHAR-M$_{LE}$}   & 8.476         & \textbf{1.968} & 6.741             &\textbf{0.897}\\
		\texttt{GeoHAR-M$_{P}$}    & 6.628         & 2.128          & 6.079             &0.982\\[5 pt]
		\texttt{Cholesky}          & 4.960         & 9.121          & \textbf{3.713}    &1.093 \\
		\texttt{GO-GARCH}          & \textbf{4.602}& 2.095          & \textbf{4.314}    &1.471 \\
		\texttt{RW}                & 7.966         & 3.077          & 6.146             &1.153  \\
		\texttt{Na\"{i}ve}         & 6.363         & 0.008          & 5.159             &0.008 \\
		\bottomrule
		\multicolumn{5}{p{0.6\textwidth}}{\footnotesize Note: The two lowest values for $\sigma_p$ and $\tau_p$ are reported in bold.}\\
	\end{tabular}
\end{table}

% \subsection{Different number of assets}

\section{Conclusions} \label{sec:Conclusions}
This study introduces a novel application of geometric deep learning to the problem of forecasting realized covariance matrices in financial markets. By leveraging the Riemannian manifold structure of SPD matrices, our approach preserves the intrinsic geometric properties of covariance matrices while avoiding the pitfalls of arbitrary parameterizations commonly used in traditional methods. The introduction of a block diagonal input matrix allows for the inclusion of multiple lagged covariance matrices, which enhances the model's ability to capture temporal dependencies in high-dimensional data and permits to extend to the matrix-variate the HAR model of \cite{cor09}.

Our empirical results, based on the S\&P 500's top 50 companies, demonstrate that the proposed method significantly improves forecasting accuracy compared to conventional approaches, including those based on Cholesky decomposition and multivariate GARCH models. In particular, a model with ten lags and trained using the logarithmic Euclidean distance as a loss function and the GeoHAR using a log-Euclidean-based Fréchet mean to compute RCOV averages perform best. Overall, our findings demonstrate that using an Euclidean loss function to train a neural network for SPD matrices leads to poor predictive results. This somehow justifies both an architecture, like the ReSPDNET, capable of properly predicting asset return covariance matrices within a geometric context, and also a geometric-aware loss metric for the training of the models.

Future research could explore further enhancements to the SPDNet architecture, such as incorporating additional non-linear activation functions different from ReLU within the RegEig layer or applying batch normalization techniques specifically designed for SPD matrices. Additionally, extending this approach to include exogenous variables in the input matrix could help understand which macroeconomic and financial variables affect the entire covariance matrix of asset returns.

\bibliographystyle{apalike}
\bibliography{manuscriptBPZ} 

\begin{thebibliography}{}

\bibitem[Andersen et~al., 2001a]{abdl01}
Andersen, T.~G., Bollerslev, T., Diebold, F.~X., and Labys, P. (2001a).
\newblock The distribution of exchange rate volatility.
\newblock {\em Journal of American Statistical Association}, 96:42--55.

\bibitem[Andersen et~al., 2001b]{Andersen2001}
Andersen, T.~G., Bollerslev, T., Diebold, F.~X., and Labys, P. (2001b).
\newblock {The distribution of realized exchange rate volatility}.
\newblock {\em The distribution of realized exchange rate volatility},
  96:42--55.

\bibitem[Andersen et~al., 2003]{Andersen2003}
Andersen, T.~G., Bollerslev, T., Diebold, F.~X., and Labys, P. (2003).
\newblock {Modeling and Forecasting Realized Volatility}.
\newblock {\em Econometrica}, 71:579--625.

\bibitem[Arsigny et~al., 2007]{Arsigny2007}
Arsigny, V., Fillard, P., Pennec, X., and Ayache, N. (2007).
\newblock {Geometric Means in a Novel Vector Space Structure on Symmetric
  Positive‐Definite Matrices}.
\newblock {\em SIAM Journal on Matrix Analysis and Applications},
  29(1):328--347.

\bibitem[Asai et~al., 2022]{Asai2022}
Asai, M., Chang, C.-L., and McAleer, M. (2022).
\newblock Realized matrix-exponential stochastic volatility with asymmetry,
  long memory and higher-moment spillovers.
\newblock {\em Journal of Econometrics}, 227(1):285--304.

\bibitem[Asai and McAleer, 2015]{Asai2015}
Asai, M. and McAleer, M. (2015).
\newblock Forecasting co-volatilities via factor models with asymmetry and long
  memory in realized covariance.
\newblock {\em Journal of Econometrics}, 189(2):251--262.

\bibitem[Baillie et~al., 1996]{Baillie1996}
Baillie, R.~T., Bollerslev, T., and Mikkelsen, H.~O. (1996).
\newblock Fractionally integrated generalized autoregressive conditional
  heteroskedasticity.
\newblock {\em Journal of Econometrics}, 74(1):3--30.

\bibitem[Barndorff-Nielsen and Shephard, 2002]{bashe02}
Barndorff-Nielsen, O.~E. and Shephard, N. (2002).
\newblock {Estimating Quadratic Variation Using Realised Variance}.
\newblock {\em Journal of Applied Econometrics}, 17:457--477.

\bibitem[Bauer and Vorkink, 2011]{bavo11}
Bauer, G.~H. and Vorkink, K. (2011).
\newblock Forecasting multivariate realized stock market volatility.
\newblock {\em Journal of Econometrics}, 160(1):93--101.

\bibitem[Bollerslev et~al., 2018]{Bollerslev2018}
Bollerslev, T., Patton, A.~J., and Quaedvlieg, R. (2018).
\newblock Modeling and forecasting (un)reliable realized covariances for more
  reliable financial decisions.
\newblock {\em Journal of Econometrics}, 207(1):71--91.

\bibitem[Brooks et~al., 2019]{brooks2019riemannian}
Brooks, D., Schwander, O., Barbaresco, F., Schneider, J.-Y., and Cord, M.
  (2019).
\newblock {Riemannian batch normalization for SPD neural networks}.
\newblock {\em Advances in Neural Information Processing Systems}, 32.

\bibitem[Bucci et~al., 2022]{bucci2022comparing}
Bucci, A., Ippoliti, L., and Valentini, P. (2022).
\newblock Comparing unconstrained parametrization methods for return covariance
  matrix prediction.
\newblock {\em Statistics and Computing}, 32(5):90.

\bibitem[Callot et~al., 2017]{Callot2017}
Callot, L. A.~F., Kock, A.~B., and Medeiros, M.~C. (2017).
\newblock Modeling and forecasting large realized covariance matrices and
  portfolio choice.
\newblock {\em Journal of Applied Econometrics}, 32(1):140--158.

\bibitem[Chakraborty et~al., 2018]{chakraborty2018statistical}
Chakraborty, R., Yang, C.-H., Zhen, X., Banerjee, M., Archer, D., Vaillancourt,
  D., Singh, V., and Vemuri, B. (2018).
\newblock A statistical recurrent model on the manifold of symmetric positive
  definite matrices.
\newblock {\em Advances in neural information processing systems}, 31.

\bibitem[Corsi, 2009]{cor09}
Corsi, F. (2009).
\newblock {A Simple Approximate Long-Memory Model of Realized Volatility}.
\newblock {\em Journal of Financial Econometrics}, 7(2):174--196.

\bibitem[DeMiguel et~al., 2009]{DeMiguel2009}
DeMiguel, V., Garlappi, L., and Uppal, R. (2009).
\newblock {Optimal Versus Naive Diversification: How Inefficient is the 1/N
  Portfolio Strategy?}
\newblock {\em The Review of Financial Studies}, 22(5):1915--1953.

\bibitem[Dryden et~al., 2009]{Dryden2009}
Dryden, I.~L., Koloydenko, A., and Zhou, D. (2009).
\newblock {Non-Euclidean Statistics for Covariance Matrices, with Applications
  to Diffusion Tensor Imaging}.
\newblock {\em The Annals of Applied Statistics}, 3(3):1102--1123.

\bibitem[Engle, 2002]{en02}
Engle, R.~F. (2002).
\newblock Dynamic conditional correlation: a simple class of multivariate
  {GARCH} models.
\newblock {\em Journal of Business and Economic Statistics}, 20:339--350.

\bibitem[Engle and Kroner, 1995]{engkro95}
Engle, R.~F. and Kroner, K.~F. (1995).
\newblock Multivariate simultaneous generalized {ARCH}.
\newblock {\em Econometric Theory}, 11:122--150.

\bibitem[Gatheral et~al., 2018]{Gatheral2018}
Gatheral, J., Jaisson, T., and Rosenbaum, M. (2018).
\newblock Volatility is rough.
\newblock {\em Quantitative Finance}, 18(6):933--949.

\bibitem[Gower, 1975]{Gower1975}
Gower, J. (1975).
\newblock {Generalized Procrustes analysis.}
\newblock {\em Psychometrika}, 40:33--50.

\bibitem[Gribisch et~al., 2020]{Gribisch2020}
Gribisch, B., Hartkopf, J.~P., and Liesenfeld, R. (2020).
\newblock Factor state--space models for high-dimensional realized covariance
  matrices of asset returns.
\newblock {\em Journal of Empirical Finance}, 55:1--20.

\bibitem[Halbleib-Chiriac and Voev, 2011]{chivo11}
Halbleib-Chiriac, R. and Voev, V. (2011).
\newblock {Modelling and Forecasting Multivariate Realized Volatility}.
\newblock {\em Journal of Applied Econometrics}, 26:922--947.

\bibitem[Han and Park, 2022]{Han2022}
Han, C. and Park, F.~C. (2022).
\newblock A geometric framework for covariance dynamics.
\newblock {\em Journal of Banking {\&} Finance}, 134:106319.

\bibitem[Hansen et~al., 2011]{haluna11}
Hansen, P.~R., Lunde, A., and Nason, J.~M. (2011).
\newblock The {Model Confidence Set}.
\newblock {\em Econometrica}, 79(2):435--497.

\bibitem[Harandi et~al., 2014]{harandi2014manifold}
Harandi, M.~T., Salzmann, M., and Hartley, R. (2014).
\newblock From manifold to manifold: Geometry-aware dimensionality reduction
  for spd matrices.
\newblock In {\em Computer Vision--ECCV 2014: 13th European Conference, Zurich,
  Switzerland, September 6-12, 2014, Proceedings, Part II 13}, pages 17--32.
  Springer.

\bibitem[Higham, 1988]{Higham1988}
Higham, N.~J. (1988).
\newblock Computing a nearest symmetric positive semidefinite matrix.
\newblock {\em Linear Algebra and its Applications}, 103:103--118.

\bibitem[Huang and Van~Gool, 2017]{huang2017riemannian}
Huang, Z. and Van~Gool, L. (2017).
\newblock {A Riemannian network for SPD matrix learning}.
\newblock In {\em Proceedings of the AAAI conference on artificial
  intelligence}, volume~31.

\bibitem[Hurvich et~al., 2005]{Hurvich2005}
Hurvich, C.~M., Moulines, E., and Soulier, P. (2005).
\newblock {Estimating Long Memory in Volatility}.
\newblock {\em Econometrica}, 73(4):1283--1328.

\bibitem[Jin and Maheu, 2016]{Jin2016}
Jin, X. and Maheu, J.~M. (2016).
\newblock Bayesian semiparametric modeling of realized covariance matrices.
\newblock {\em Journal of Econometrics}, 192(1):19--39.

\bibitem[Lanne and Saikkonen, 2007]{Lanne2007}
Lanne, M. and Saikkonen, P. (2007).
\newblock {A Multivariate Generalized Orthogonal Factor GARCH Model}.
\newblock {\em Journal of Business {\&} Economic Statistics}, 25(1):61--75.

\bibitem[Laurent et~al., 2013]{Laurent2013}
Laurent, S., Rombouts, J.~V., and Violante, F. (2013).
\newblock On loss functions and ranking forecasting performances of
  multivariate volatility models.
\newblock {\em Journal of Econometrics}, 173(1):1--10.

\bibitem[Lee, 2024]{Lee2024}
Lee, H.-T. (2024).
\newblock Riemannian-geometric regime-switching covariance hedging.
\newblock {\em Journal of Futures Markets}, 44(6):1003--1054.

\bibitem[Li et~al., 2012]{li2012electroencephalogram}
Li, Y., Wong, K.~M., and de~Bruin, H. (2012).
\newblock {Electroencephalogram signals classification for sleep-state
  decision--a Riemannian geometry approach}.
\newblock {\em IET signal processing}, 6(4):288--299.

\bibitem[Markowitz, 1952]{ma52}
Markowitz, H. (1952).
\newblock Portfolio selection.
\newblock {\em Journal of Finance}, 7(1):77--91.

\bibitem[Marron and Dryden, 2022]{marron2022}
Marron, J. and Dryden, I. (2022).
\newblock {\em {Object Oriented Data Analysis}}.
\newblock Chapman \& Hall.

\bibitem[Noureldin et~al., 2011]{nosheshe11}
Noureldin, D., Shephard, N., and Sheppard, K. (2011).
\newblock Multivariate high-frequency-based volatility ({HEAVY}) models.
\newblock {\em Journal of Applied Econometrics}, 27(6):907--933.

\bibitem[Opschoor et~al., 2024]{Opschoor2024}
Opschoor, A., Lucas, A., and Rossini, L. (2024).
\newblock {The Conditional Autoregressive F-Riesz Model for Realized Covariance
  Matrices}.
\newblock {\em Journal of Financial Econometrics}.

\bibitem[Pascalau and Poirier, 2023]{Pascalau2023}
Pascalau, R. and Poirier, R. (2023).
\newblock Increasing the information content of realized volatility forecasts*.
\newblock {\em Journal of Financial Econometrics}, 21(4):1064--1098.

\bibitem[Reiss and Winkelmann, 2021]{Reiss2021}
Reiss, M. and Winkelmann, L. (2021).
\newblock Inference on the maximal rank of time-varying covariance matrices
  using high-frequency data.

\bibitem[Schwert, 1989]{sch89}
Schwert, W.~G. (1989).
\newblock Why does stock market volatility change over time?
\newblock {\em Journal of Finance}, 44(5):1115--1153.

\bibitem[Tuzel et~al., 2008]{tuzel2008pedestrian}
Tuzel, O., Porikli, F., and Meer, P. (2008).
\newblock {Pedestrian detection via classification on Riemannian manifolds}.
\newblock {\em IEEE transactions on pattern analysis and machine intelligence},
  30(10):1713--1727.

\bibitem[van~der Weide, 2002]{vanderWeide2002}
van~der Weide, R. (2002).
\newblock {GO-GARCH: A Multivariate Generalized Orthogonal GARCH Model}.
\newblock {\em Journal of Applied Econometrics}, 17(5):549--564.

\bibitem[Vrontos et~al., 2003]{Vrontos2003}
Vrontos, I.~D., Dellaportas, P., and Politis, D.~N. (2003).
\newblock {A full-factor multivariate GARCH model}.
\newblock {\em The Econometrics Journal}, 6(2):312--334.
\newblock Full publication date: 2003.

\bibitem[Wang et~al., 2022]{wang2022dreamnet}
Wang, R., Wu, X.-J., Chen, Z., Xu, T., and Kittler, J. (2022).
\newblock {Dreamnet: A deep Riemannian manifold network for spd matrix
  learning}.
\newblock In {\em Proceedings of the Asian Conference on Computer Vision},
  pages 3241--3257.

\bibitem[Yu et~al., 2017]{Yu2017}
Yu, P. L.~H., Li, W.~K., and Ng, F.~C. (2017).
\newblock {The Generalized Conditional Autoregressive Wishart Model for
  Multivariate Realized Volatility}.
\newblock {\em Journal of Business {\&} Economic Statistics}, 35(4):513--527.

\bibitem[Zhang et~al., 2024]{Zhang2024}
Zhang, C., Pu, X., Cucuringu, M., and Dong, X. (2024).
\newblock {Graph-Based Methods for Forecasting Realized Covariances}.
\newblock {\em Journal of Financial Econometrics}, page nbae026.

\bibitem[Zhao et~al., 2023]{zhao2023machine}
Zhao, C., Dong, E., Tong, J., Yang, S., and Du, S. (2023).
\newblock {Machine Learning Classification of Riemannian Tangent Spaces Based
  on MI-BCI}.
\newblock In {\em 2023 IEEE International Conference on Mechatronics and
  Automation (ICMA)}, pages 807--812. IEEE.

\end{thebibliography}

\end{document}